
\input harvmac.tex
\input epsf.tex
\def\rb{\bar{\rho}}
\def\rbt{\tilde{\rb}}
\def\rt{\tilde{\rho}}

\def\zdag{Z^\dagger}
\def\emrvac{|\emptyset \rangle }
\def\emlvac{\langle \emptyset |}

\def\alt{\tilde{\al}}

\def\mass{{\rm m }}
\def\ma{\mass}
\def\ha{ {\scriptstyle{\inv{2}} }}
\def\tha{ {\scriptstyle{\frac{3}{2}} }}

\def\dag{\dagger}
\def\betah{{\hat{\beta}}}
\def\hp{\CH_\CP}
\def\hf{\CH_\CF}

\def\slh{{\hat{sl(2)}}}
\def\psib{\bar{\psi}}
\def\psip{\psi^+}
\def\psim{\psi^-}
\def\psibp{\psib^+}
\def\psibm{\psib^-}

\def\vphi{\varphi}
\def\bh{\hat{b}}
\def\bhp{\bh^+}
\def\bhm{\bh^-}
\def\sqm{\sqrt{\ma}}
\def\ez{e^{\ma z u + \ma \zb /u}}
\def\squ{\sqrt{u}}

\def\bb{\bar{b}}
\def\bbp{\bar{b}^+}
\def\bbm{\bar{b}^-}
\def\du{ \frac{du}{2 \pi i |u|} }
\def\dua{ \frac{du}{2 \pi i u} }

\def\lvacp{\langle +\ha \vert}
\def\lvacm{\langle -\ha \vert}

\def\lvacpm{\langle \pm \ha \vert}
\def\rvacpm{\vert \pm \ha \rangle}
\def\rvacmp{\vert \mp \ha \rangle}
\def\rvacp{\vert + \ha \rangle}
\def\rvacm{\vert - \ha \rangle}
\def\va#1{\vert {#1} \rangle}
\def\lva#1{\langle {#1} \vert}

\def\dstate{ {}^{\ep_1 \cdots \ep_n} \lva{\th_1 \cdots \th_n} }
\def\tstate{ {}^{+-} \lva{\th_1 , \th_2 } }

\def\col#1#2{\left(\matrix{#1\cr #2\cr}\right)}

\def\psipm{\psi^\pm}
\def\psibpm{\psib^\pm }
\def\bhpm{\bh^\pm}
\def\bpm{b^\pm}
\def\bbpm{\bar{b}^\pm}

\def\bhpm{\bh^\pm}
\def\cL{\CC^{L}_{\vphi}}
\def\cR{\CC^{R}_{\vphi} }

%
%
%
%

\def\tilde{\widetilde}
\def\bar{\overline}
\def\hat{\widehat}
\def\*{\star}
\def\[{\left[}
\def\]{\right]}
\def\({\left(}		
\def\){\right)}

%
%
\def\zb{{\bar{z} }}
\def\frac#1#2{{#1 \over #2}}
\def\inv#1{{1 \over #1}}

\def\d{\partial}

\def\rvac{\hbox{$\vert 0\rangle$}}
\def\lvac{\hbox{$\langle 0 \vert $}}
\def\2pi{\hbox{$2\pi i$}}

\def\dsl{\raise.15ex\hbox{/}\kern-.57em\partial}
\def\Dsl{\,\raise.15ex\hbox{/}\mkern-.13.5mu D}
%
%
\def\th{\theta}		
		\def\Ga{\Gamma}

\def\al{\alpha}
\def\ep{\epsilon}
\def\la{\lambda}	
\def\de{\delta}		
\def\om{\omega}		
	
\def\vphi{\varphi}
%
%
		\def\CC{{\cal C}}
		\def\CF{{\cal F}}
	\def\CH{{\cal H}}	
		
		\def\CO{{\cal O}}
\def\CP{{\cal P}}		
		\def\CU{{\cal U}}

\def\rvac{\hbox{$\vert 0\rangle$}}
\def\lvac{\hbox{$\langle 0 \vert $}}

\def\2pi{\hbox{$2\pi i$}}

\def\dsl{\raise.15ex\hbox{/}\kern-.57em\partial}
\def\Dsl{\,\raise.15ex\hbox{/}\mkern-.13.5mu D}
%
%
%
\font\numbers=cmss12
\font\upright=cmu10 scaled\magstep1
\def\stroke{\vrule height8pt width0.4pt depth-0.1pt}
\def\topfleck{\vrule height8pt width0.5pt depth-5.9pt}
\def\botfleck{\vrule height2pt width0.5pt depth0.1pt}
\def\Zmath{\vcenter{\hbox{\numbers\rlap{\rlap{Z}\kern
0.8pt\topfleck}\kern
2.2pt
                   \rlap Z\kern 6pt\botfleck\kern 1pt}}}
\def\Qmath{\vcenter{\hbox{\upright\rlap{\rlap{Q}\kern
                   3.8pt\stroke}\phantom{Q}}}}
\def\Nmath{\vcenter{\hbox{\upright\rlap{I}\kern 1.7pt N}}}
\def\Cmath{\vcenter{\hbox{\upright\rlap{\rlap{C}\kern
                   3.8pt\stroke}\phantom{C}}}}
\def\Rmath{\vcenter{\hbox{\upright\rlap{I}\kern 1.7pt R}}}
\def\Z{\ifmmode\Zmath\else$\Zmath$\fi}
\def\Q{\ifmmode\Qmath\else$\Qmath$\fi}
\def\N{\ifmmode\Nmath\else$\Nmath$\fi}
\def\C{\ifmmode\Cmath\else$\Cmath$\fi}
\def\R{\ifmmode\Rmath\else$\Rmath$\fi}
\def\Zmath{Z}

\def\AASS{E. Abdalla, M. C. B. Abdalla, G. Sotkov, and M. Stanishkov,
{\it Off Critical Current Algebras}, Univ. Sao Paulo preprint,
IFUSP-preprint-1027, Jan. 1993.}
\def\Ginsparg{P. Ginsparg, Les Houches Lectures 1988, in
{\it Fields, Strings and Critical Phenomena}, E. Br\'ezin and J. Zinn-Justin
Eds., North Holland (1990).}
\def\Griffin{P. Griffin, Nucl. Phys. B334, 637.}
\def\MSS{B. Schroer and T. T. Truong, Nucl. Phys. B144 (1978) 80 \semi
E. C. Marino, B. Schroer, and J. A. Swieca, Nucl. Phys. B200 (1982) 473.}
\def\FrResh{I. B. Frenkel and N. Yu. Reshetikhin, Commun. Math. Phys. 146
(1992) 1. }
\def\VVstar{B. Davies, O. Foda, M. Jimbo, T. Miwa and A. Nakayashiki,
Commun. Math. Phys. 151 (1993) 89;
M. Jimbo, K. Miki, T. Miwa and A. Nakayashiki,
Phys. Lett. A168 (1992) 256.}
\def\BLnlc{D. Bernard and A. LeClair, Commun. Math. Phys. 142 (1991) 99;
Phys. Lett. B247 (1990) 309.}
\def\form{F. A. Smirnov, {\it Form Factors in Completely Integrable
Models of Quantum Field Theory}, in {\it Advanced Series in Mathematical
Physics} 14, World Scientific, 1992.}
\def\BPZ{A. A. Belavin, A. M. Polyakov, and A. B. Zamolodchikov,
Nucl. Phys. B241 (1984) 333.}
\def\Frenkeli{I. B. Frenkel and N. Jing,
Proc. Natl. Acad. Sci. USA 85 (1988) 9373.}
\def\ZZ{A. B. Zamolodchikov and Al. B. Zamolodchikov, Annals
Phys. 120 (1979) 253.}
\def\Colemani{S. Coleman, Phys. Rev. D 11 (1975) 2088.}
\def\TI{H. Itoyama and H. B. Thacker, Nucl. Phys. B320 (1989) 541.}
%

\Title{CLNS 93/1263}
{\vbox{\centerline{ Particle-Field Duality    }
\centerline{ and } \centerline{Form Factors from Vertex Operators} }}

\bigskip
\bigskip

\centerline{Costas Efthimiou and Andr\'e LeClair}
\medskip\centerline{Newman Laboratory}
\centerline{Cornell University}
\centerline{Ithaca, NY  14853}
\bigskip\bigskip

\vskip .3in

Using a duality between the space of particles and the space
of fields, we show how one can compute form factors directly
in the space of fields.  This introduces the notion of vertex
operators, and form factors are vacuum expectation values of
such vertex operators in the space of fields.  The vertex operators
can be constructed explicitly in radial quantization.  Furthermore,
these vertex operators can be exactly bosonized in momentum space.
We develop these ideas by studying the free-fermion point of the
sine-Gordon theory, and use this scheme to compute some form-factors
of some non-free fields in the sine-Gordon theory.  This work
further clarifies earlier work of one of the authors, and extends
it to include the periodic sector.

\Date{12/93}
%
%
%
%
%
\noblackbox
\def\ot{\otimes}
\def\zb{{\bar{z}}}

\def\zbar{{\bar{z}}}
%
%
%
%
%
%
%
%
%
%

\newsec{Introduction}

For relativistic quantum field theories with a massive particle
spectrum, the main dynamical properties one is interested in are
the S-matrix, the form factors of all local fields, and the
Green's functions of these fields.   For the integrable quantum
field theories in 2 space-time dimensions, some of these properties
have been computed exactly.  The algebraic structures that
characterize the S-matrices are well-known \ref\rzz{\ZZ}, and
have been used to compute them for a wide variety of models.
Bootstrap axioms satisfied by the form factors have been
formulated\ref\rkw{M. Karowski and P. Weisz, Nucl. Phys. B139 (1978)
445.}\ref\rform{\form}.  Important progress in solving the
bootstrap for the multiparticle form factors was made by
Smirnov\ref\rffold{F. A. Smirnov, J. Phys. A: Math. Gen. 19 (1986)
L575.}\rform, where he computed exactly the form factors of
certain basic fields,  such as the energy-momentum tensor and
global conserved currents, in the sine-Gordon (SG) model,
$SU(N)$ Thirring model, and $O(3)$ non-linear sigma model.

It is of interest to develop a more algebraic framework for the
computation of form factors, with the aim of constructing
the solutions for the complete set of fields in an
efficient manner.  A deeper understanding of such algebraic
structures is likely to facilitate generalizations to other
models, and could lead to some much-needed new
approaches to the problem of computing Green's functions.
In the works
\ref\lec{A. LeClair, {\it Spectrum Generating Affine Lie Algebras
in Massive Field Theory}, hep-th/9305110, to appear in Nucl. Phys. B.}
\ref\rluk{S. Lukyanov, {\it Free Field Representation for Massive
Integrable Models}, and {\it Correlators of the Jost
Functions in the Sine-Gordon Model}, Rutgers preprints
RU-93-30, RU-93-55.}
two new approaches to the computation of form factors were proposed.
In \lec, structures in radial quantization were used to construct
form factors explicitly as vacuum expectations of vertex operators
in momentum space.
The ideas underlying the construction were developed at the
free fermion point of the sine-Gordon theory.  This case is not
completely trivial, since there are fields in the sine-Gordon
theory, such as $\exp (\pm i \phi /2 )$, with non-trivial form factors
and Green's functions since these fields are non-local in terms of
the fermion fields.
In Lukyanov's approach\rluk, the form factors are constructed as
traces over auxiliary Fock-modules.
The original motivation behind his construction came
from the work\ref\vvstar{\VVstar}, where the necessary
properties of these traces were understood in the
context of lattice models.  The significant differences between
the approaches in \lec\ and \rluk\ is explained by the fact that
whereas in \lec\ {\it radial} quantization is used,
the construction in \rluk\ appears to correspond to {\it angular}
quantization.
To see this,
define the usual Euclidean light-cone and polar coordinates  as
follows
\eqn\coor{
z = (t+ix)/2 =\frac{r}{2} \exp(i\vphi) , ~~~~~~~~
\zbar = (t-ix)/2 = \frac{r}{2} \exp (-i\vphi ) . }
In radial quantization $r$ is declared as the `time',
whereas in angular quantization $\vphi$ is the `time'.
In angular quantization, since the Lorentz boost operator $L$
generates shifts in $\vphi$, it is the Hamiltonian.  Then,
functional integrals can be represented as traces:
\eqn\func{
\lvac ~ \CO ~ \rvac  = \frac{\int ~ D\Phi  ~ e^{-S} ~ \CO}
{\int D\Phi e^{-S} }
{}~ = ~  \frac{Tr ~ ( e^{2\pi i L} \> \CO )}{Tr ~ ( e^{2\pi i L} )} . }
The $2\pi i$ constant in the factor $\exp (2\pi i L )$
is fixed by the
$2\pi$ length of the `time' $\vphi$.  In \vvstar\rluk,
the latter constant was fixed by imposing  the right
symmetry properties of the form factors expressed as these
traces.

In this paper, we elaborate further on the construction in \lec.
We explain quite generally how the computation of form factors
can be carried out in the space of fields, and how this is dual
to the usual computation in the space of particles.  This leads
to a precise notion of vertex operators and their utility for
computing form factors.
We continue to study the free fermion point of the sine-Gordon
theory, since our interests at this stage are mainly to understand
more completely the physical features that make this new approach
possible.
With this understanding, we treat also
the periodic sector, which was not done in \lec.

\newsec{Particle-Field Duality and Form-Factors}

In a quantum field theory with a spectrum of massive particles,
one deals with the space of multiparticle states $\hp$.
In the context of the integrable theories in 2 dimensions,
we can describe $\hp$ as follows.  Introduce the so-called
Faddeev-Zamolodchikov operators $\zdag_\ep (\th ) ,
Z^\ep (\th )$, where $\th$ is the rapidity parameterizing
the energy-momentum ($E= \ma \cosh \th ,~ P = \ma \sinh \th $),
and $\ep$ is an isotopic index, satisfying:
\eqn\eIIi{\eqalign{
Z^{\ep_1} (\th_1 ) \> Z^{\ep_2} (\th_2 )
&= S^{\ep_1 \ep_2}_{\ep'_1 \ep'_2} (\th_{12} ) ~
Z^{\ep'_2} (\th_2 ) \> Z^{\ep'_1} (\th_1 )
\cr
\zdag_{\ep_1} (\th_1 ) \> \zdag_{\ep_2} (\th_2 )
&= S^{\ep'_1 \ep'_2}_{\ep_1 \ep_2} (\th_{12} ) ~
\zdag_{\ep'_2} (\th_2 ) \> \zdag_{\ep'_1} (\th_1 )
\cr
Z^{\ep_1} (\th_1 ) \zdag_{\ep_2} (\th_2 ) ~&- ~
S^{\ep'_2 \ep_1}_{\ep_2 \ep'_1} (\th_{21} ) ~
\zdag_{\ep'_2} (\th_2) Z^{\ep'_1} (\th_1 )
= \delta^{\ep_1}_{\ep_2} ~  \delta (\th_1 - \th_2 ) . \cr }}
Above, $S$ is the S-matrix, and $\th_{12} = \th_1 - \th_2$.
 Define particle states as follows:
\eqn\eIIii{\eqalign{
\va{\th_1 \cdots \th_n }_{\ep_1 \cdots \ep_n }
&= \zdag_{\ep_1} (\th_1 ) \cdots \zdag_{\ep_n} (\th_n ) \rvac
\cr
{}^{\ep_1 \cdots \ep_n } \lva{\th_1 \cdots \th_n}
&= \lvac Z^{\ep_1} (\th_1) \cdots Z^{\ep_n} (\th_n ) . \cr }}
The space $\hp$ and its dual $\hp^*$ are spanned by the above
states:
$$\hp = \{ \oplus_n ~
\va{\th_1 \cdots \th_n }_{\ep_1 \cdots \ep_n }
\}$$
$$\hp^* = \{ \oplus_n ~
{}^{\ep_1 \cdots \ep_n } \lva{\th_1 \cdots \th_n}
\}.$$
In this space one has the completeness relation
\eqn\eIIiv{
1 = \sum_{\vec{\th}} \va{\overrightarrow{\th}} \lva{\overleftarrow{\th}}
=
\sum_{n=0}^\infty \inv{n!} \sum_{ \{ \ep_i \} } \int d\th_1 \cdots
d\th_n ~ |\th_1 ,\cdots , \th_n \rangle_{\ep_1 \cdots \ep_n}
{}^{\ep_n \cdots \ep_1 } \langle \th_n , \cdots , \th_1 | . }

Next consider the space of fields $\hf$.  Let $\CF$ denote the
complete space of fields, and define $\va{\Phi_i} = \Phi_i (0) \rvac$,
$\Phi_i (x)  \in \CF$.
The space $\hf$ is defined as follows
\eqn\eIIv{
\hf = \{ \oplus_{\Phi_i \in \CF}  ~~ \va{\Phi_i}  \} .}
Form factors are matrix elements of fields in the space of states
$\hp$.  The basic form factors
$\dstate\Phi \rangle$, from which the more general matrix
elements may be obtained by  crossing symmetry, are inner products
of states in $\hf$ with states in $\hp^*$.  The completeness relation
\eIIiv\ allows us to map states in $\hf$ to states in $\hp$, i.e. to
view $\va{\Phi} \in \hp$:
\eqn\eIIvi{
\va{\Phi_i} =
 \sum_{\vec{\th}} \va{\overrightarrow{\th}}
 \lva{\overleftarrow{\th}} \Phi_i \rangle .}
The intuitive simplicity of the space $\hp$ is responsible for this
conventional way of thinking about form factors.

We give now a dual description of the same form factors.  Let us
suppose that one can define a dual to the space of fields $\hf^*$
with inner product and completeness relation:
\eqn\eIIvii{\eqalign{
\lva{\Phi^i} \Phi_j \rangle &= \delta^i_j  \cr
1 &= \sum_i \va{\Phi_i} \lva{\Phi^i }  . \cr }}
Then one can map a state $\va{\vec{\th}} \in \hp$ into
$\hf$.  The dual statement is
\eqn\eIIviii{
\dstate = \sum_{\Phi_i \in \CF}
\dstate \Phi_i \rangle \lva{\Phi^i } ~~~~\in \hf^* . }

In order to work the above simple remarks into an efficient means
of computing form factors, one must work explicitly with the space
$\hf$.  The space $\hf$ diagonalizes the Lorentz boost operator $L$,
thus it can be understood as the space of radial quantization
\ref\rfhj{S. Fubini, A. J. Hanson, and
R. Jackiw, Phys. Rev. D7 (1973) 1732.}\lec.  We remark that in conformal
field theory\ref\rbpz{\BPZ}\ one deals precisely with the space
of fields, and it was argued in \lec\ that for many massive quantum
field theories, the space $\hf$ is identical to its description
in the ultraviolet conformal field theory.

In order to use these ideas to compute form factors, we need to
introduce the notion of vertex operators.  The formula
\eIIviii\ implies that one can map states in $\hp^*$ to states
in $\hf^*$.  We call this map the `particle-field map'.
We construct this map explicitly by defining vertex
operators  $V^\ep (\th )$ as follows:
\eqn\eIIix{
\dstate = \lva{\Omega} ~ V^{\ep_1} (\th_1 ) \cdots
V^{\ep_n} (\th_n ) ~~ \in \hf^* }
where $\lva{\Omega}$ is a fixed `vacuum' state. The
vertex operators are distinguished from the Faddeev-Zamolodchikov
operators $Z(\th )$ since they act on completely different spaces.
However the basic algebraic relations satisfied by the $Z$ operators
continue to be satisfied by the $V$ operators.
The vertex operators $V^{\ep} (\th )$ operate in the space $\hf$:
\eqn\eIIx{
V^{\ep} (\th ) : ~~~~\hf \to \hf . }
In the sequel we will describe how to construct these vertex
operators explicitly using radial quantization.  Once the
vertex operators are constructed, the form factors
$\dstate \Phi_i \rangle$ are computed directly in the
discrete space $\hf$ using \eIIix.

We close this section with some remarks on Green's functions.
One way to define Green's functions is to insert the identity
\eIIiv\ between fields in the matrix element, leading to an
infinite integral representation.  Just as for form factors, one
can instead insert the identity in $\hf$.  The fundamental object
of interest is the following matrix element:
\eqn\eIIope{
C_{ij}^k (x) = \lva{\Phi^k} \> \Phi_i (x) \> \va{\Phi_j} .}
These are equivalent to the operator product coefficients:
\eqn\eIIopeb{
\Phi_i (x) \Phi_j (0)  = \sum_k ~ C_{ij}^k (x) ~ \Phi_k (0) . }
Defining $\Phi_0 \equiv 1$, one sees that
$C^0_{ij} (x) = \lvac \Phi_i (x) \Phi_j (0) \rvac $ are the
2-point functions.  Multi-point functions can be expressed in terms
of the above coefficients.  Manifest translation invariance is
lost in the resulting expressions, as one would expect from
radial quantization.  Though these structures provide a different
point of view on the correlation functions, by themselves they do
not appear to simplify the problem.

\newsec{Radial Quantization}

We will consider the sine-Gordon theory, defined by the action
\eqn\eIi{
S = \inv{4\pi} \int d^2 z \(  \d_z \phi \d_\zb \phi
                 + 4\la  \cos ( \betah \phi ) \) , }
at the free fermion point, which occurs at
$\hat{\beta} = 1$\ref\rcol{\Colemani}.
The free Dirac fermion fields $\psi^\pm $,
$\psib^\pm$ carry $U(1)$ charge $\pm 1$, and their dynamics
is governed by the action
\eqn\eIIi{
S = - \inv{4\pi} \int dx dt \(
\psibm \d_z \psibp + \psim \d_\zb \psip
+ i \ma ( \psim \psibp - \psibm \psip ) \) , }
with the equations of motion
\eqn\eIIIii{
\d_z \psib^\pm = i \ma \psi^\pm ,~~~~~
\d_\zb \psi^\pm = -i \ma \psib^\pm . }

In conventional temporal quantization, the expansion of
the fermion fields in terms of momentum space operators is
as follows:
\eqn\eIIxvib{\eqalign{
\psip (x,t) &=  \sqm \int_{-\infty}^\infty
d\th ~ e^{\th/2}
\( Z^+ (\th ) \> e^{-ip(\th ) \cdot x }  - \zdag_- (\th )
\> e^{i p (\th)  \cdot x } \)
\cr
\psibp (x,t) &=  -i \sqm \int_{-\infty}^\infty
d\th ~ e^{-\th/2}
\( Z^+ (\th ) \> e^{-ip(\th ) \cdot x }  + \zdag_- (\th ) \> e^{i p(\th )
\cdot x } \)
,\cr}}
with $\psim = (\psi^+)^\dag , \psib^- = (\psib^+)^\dag$, and
$(Z^\ep )^\dagger = Z^\dag_\ep $.

Consider now radial quantization of the free-fermion
theory\rfhj\ref\rit{\TI}\ref\rgrif{\Griffin}.  Define
the usual polar coordinates as in \coor.
One can define
two distinct sectors, the periodic (p) and
anti-periodic (a), with expansions
\eqn\eIIIxvii{
\Psi_{(a,p)}^\pm = \col{\psibpm}{\psipm} = \sum_\om
 \left[ \bpm_\om ~ \Psi^{(a,p)}_{-\om - 1/2} ~+~
\bbpm_\om ~ \bar{\Psi}^{(a,p)}_{-\om -1/2}  \right] , }
where for the periodic sector $\om \in \Zmath + 1/2$, and
for the anti-periodic sector $\om \in \Zmath$.
The basis spinors can be found as solutions to the
equations of motion \eIIIii\ in radial coordinates:
\eqn\eIIIxviii{\eqalign{
\Psi^{(a)}_{-\om -1/2} &=
\Gamma (\ha - \om ) ~ \ma^{\om + 1/2} ~
\col{i \> e^{i(\ha - \om )\vphi} ~ I_{\ha - \om} (\ma r) }
{e^{-i(\om + \ha )\vphi} ~ I_{-\om -\ha} (\ma r)}  \cr
\bar{\Psi}^{(a)}_{-\om - 1/2}
&=
\Gamma (\ha - \om ) ~ \ma^{\om + 1/2} ~
\col{ e^{i(\ha + \om )\vphi} ~ I_{-\ha - \om} (\ma r) }
{-i \> e^{-i(\ha - \om )\vphi} ~ I_{\ha - \om } (\ma r)}  ,\cr}}
\eqn\eIIIxix{\eqalign{
\cr
\Psi^{(p)}_{-\om -1/2} &= \frac{2 \ma^{\om + 1/2}}{\Ga (\ha + \om )}
{}~ \col{-i \> e^{i(\ha - \om )\vphi} ~ K_{ \om - \ha} (\ma r) }
{e^{-i(\om + \ha)\vphi} ~ K_{\om +\ha} (\ma r)} ~~~~\om \geq 1/2   \cr
\bar{\Psi}^{(p)}_{-\om -1/2} &= \frac{2 \ma^{\om + 1/2}}{\Ga (\ha + \om )}
{}~ \col{ e^{i(\ha + \om )\vphi} ~ K_{ \om + \ha} (\ma r) }
{i \> e^{i(\om - \ha)\vphi} ~ K_{\om -\ha} (\ma r)} ~~~~\om \geq 1/2
, \cr}}
and $\Psi^{(p)}_{-\om - 1/2}$, $\bar{\Psi}^{(p)}_{-\om -1/2} $
for $\om \leq -1/2$ have the same expression as in the anti-periodic sector.

In the quantum theory these modes satisfy simple anti-commutation
relations:
\eqn\eIIIxxiv{
\{ b^+_\om , b^-_{\om'} \} =
\{ \bbp_\om , \bbm_{\om'} \} = \de_{\om , -\om'}
, ~~~~~\{ b_\om , \bb_{\om '} \} = 0. }
These modes diagonalize the Lorentz boost operator:
\eqn\eIIIxxiiib{
\[ L , \bpm_\om \] = -\om ~ \bpm_\om , ~~~~~
\[ L , \bbpm_\om \] = \om ~ \bbpm_\om . }
The space of radial quantization consists of Fock modules built
from these oscillators.  In the periodic sector, the vacuum is
defined to satisfy
\eqn\eIIIxxv{
\bpm_\om \> \rvac = \bbpm_\om \> \rvac = 0 , ~~~~~\om \geq 1/2 . }
The vacuum $\rvac$ is the physical one.  One constructs
Fock modules for the periodic sector as follows:
\eqn\eIIIxxvi{
\CH^L_p = \left\{ b^{\ep_1}_{-\om_1}
b^{\ep_2}_{-\om_2} \cdots
 \rvac \right\} , ~~~~~~~
\CH^R_p = \left\{ \bb^{\ep_1}_{-\om_1}
\bb^{\ep_2}_{-\om_2} \cdots
 \rvac \right\} , }
where $\om_i \in \Zmath + 1/2 , \om_i  \geq 1/2$.

In the anti-periodic sector, due to the existence of the
zero modes, the `vacuum' states are doubly degenerate.
These vacua $\va{\pm \ha}_L$ and $\va{\pm \ha}_R$ are
characterized as follows:
\eqn\eIIIxxxiv{\eqalign{
\bpm_0 \> \va{\mp \ha}_L = \rvacpm_L , ~~~~ \bpm_0 \> \rvacpm_L = 0 ,
{}~~~~ \bpm_n \> \rvacpm_L = 0,~~~~&n\geq 1 \cr
\bbpm_0 \> \va{\mp \ha}_R = \rvacpm_R , ~~~~ \bbpm_0 \> \rvacpm_R = 0 ,
{}~~~~ \bbpm_n \> \rvacpm_R = 0,~~~~&n\geq 1 . \cr}}
The dual vacua $\lvacpm$ are defined by the inner products
\eqn\eIIIxxxv{
{}_L \lva{\mp \ha} \pm \ha \rangle_L = {}_R \langle \mp \ha \rvacpm_R = 1.}
The anti-periodic
Fock spaces are defined as
\eqn\eIIIxxxvb{
\CH^L_{a_{\pm}}
= \left\{ b^{\ep_1}_{-n_1} b^{\ep_2}_{-n_2} \cdots
\cdots \rvacpm_L \right\} , ~~~~~
\CH^R_{a_{\pm}}
= \left\{ \bb^{\ep_1}_{-n_1} \bb^{\ep_2}_{-n_2} \cdots
\cdots \rvacpm_R \right\} , }
for $n_i \in \Zmath  \geq 1$.

The space of radial quantization corresponds precisely to the
space of fields $\hf$ described in the last section.
Namely, in the periodic sector $\hf^{(p)} = \CH^L_p \otimes
\CH^R_p$, and in the
anti-periodic sector $\hf^{(a)} = \CH^L_a \ot \CH^R_a$,
where
$\CH^{L,R}_a = \CH^{L,R}_{a_+} \oplus \CH^{L,R}_{a_-} $.
What is remarkable about this result is that the structure
of the space of fields is identical to that in the massless
conformal limit.  One finds that in the limit $r\to 0$,
the Bessel functions in the expansions \eIIIxvii\ behave in
such a way that the mass-dependent terms in the
expression $\Phi (r) \rvac$ disappear.
Arguments explaining this phenomenon were given in \lec.

We present some simple examples that we will use later.
The $U(1)$ current $J_\mu$ has components
$J_z = \psi^+ \psi^-$, $J_\zbar = \psib^+ \psib^- $.
Also of interest is the energy-momentum tensor:\foot{We
normalize conserved currents such that
$Q = \inv{4\pi} \int dx \( J_z + J_\zbar \) $ is the properly
normalized conserved charge.}
\eqn\eIIIxiii{
\eqalign{
T_{zz} &= \inv{2} \( \psi^- \d_z \psi^+  - \d_z \psi^- \psi ^+ \)
, ~~~~~
T_{\zbar\zbar} = \inv{2} \( \psib^- \d_\zbar \psib^+
- \d_\zbar \psib^- \psib^+ \)  \cr
T_{z\zbar} &= \inv{2} \( \psi^- \d_\zbar \psi^+  - \d_\zbar \psi^- \psi^+ \)
, ~~~~~
T_{\zbar z} = \inv{2} \( \psib^- \d_z \psib^+ - \d_z \psib^- \psib^+ \)
. \cr }}
Using the expansions \eIIIxvii, the asymptotic expansions of the
Bessel functions as $r\to 0$, and the properties
of the vacuum \eIIIxxv, one finds
\eqn\eIIIxiv{\eqalign{
\d_z^n \psi^\pm (0) \rvac &= n! \> b^\pm_{-n-\ha} \rvac ,
{}~~~~~~\d_\zbar^n \psib^\pm (0) \rvac = n! \bb^\pm_{-n-\ha} \rvac \cr
J_z (0) \rvac &= b^+_{-\ha} b^-_{-\ha} \rvac , ~~~~~~
J_\zbar (0) \rvac = \bb^+_{-\ha} \bb^-_{-\ha} \rvac \cr
T_{zz} (0) \rvac &= \inv{2} \( b^-_{-\ha} b^+_{-\tha}
- b^-_{-\tha} b^+_{-\ha} \) \rvac , ~~~~~
T_{\zbar\zbar} (0) \rvac
= \inv{2} \( \bb^-_{-\ha} \bb^+_{-\tha}
- \bb^-_{-\tha} \bb^+_{-\ha} \) \rvac \cr
T_{z \zbar} (0)\rvac &= T_{\zbar z} (0) \rvac
= - \frac{i\ma}{2} \( b^-_{-\ha} \bb^+_{-\ha}  - \bb^-_{-\ha} b^+_{-\ha}
\) \rvac . \cr}}

The anti-periodic sector is more interesting, since here
one can access fields in the sine-Gordon theory which are
not simply expressed in terms of the fermion fields.  By studying
the operator product expansion, the following identification
was made in \lec:
\eqn\eIIIxvi{
e^{\pm i \phi (0)/2 } \> \rvac ~=~
(c\ma )^{1/4} ~ \( \rvacpm_L \ot \rvacmp_R \)
\equiv (c\ma )^{1/4} ~ \va{\pm \ha}
 , }
where $\phi (z, \zb )$ is the local SG field, and
$c$ is an undetermined dimensionless constant.
The mass dimension of $1/4$ on the RHS is fixed by
the known scaling dimension of the fields $\exp (\pm i \phi /2 )$,
which is the same as in the conformal limit.
(The constant $c$ can
be fixed by specifying the 1-point functions of these fields;
see below.)   The fields $\exp( \pm i \phi /2 )$ are non-trivial
terms of the fermions, since the bosonization
relation is $\cos( \phi ) = ( \psi^- \psib^+ - \psib^- \psi^+ )$.
All other states in $\hf^{(a)}$ correspond to regularized products of fermion
fields and their derivatives with the basic fields
$\exp (\pm i \phi /2 )$.

\newsec{Particle-Field Maps}

The basic property that allows us to construct explicitly the
particle-field maps of the form \eIIix\ is the fact that the
space of radial quantization can be obtained from the usual
space of temporal quantization by appropriate analytic continuation
in momentum space.  Since the situation in the periodic versus
the anti-periodic sector is quite different, we treat them separately.

\medskip
\noindent
4.1 {\it Periodic Sector}

Let us combine the creation and annihilation operators of temporal
quantization into a single operator as follows.  Define the
momentum space variable $u$ as
\eqn\eIVi{
u = e^\th }
and define operators $\bh^\pm (u) $ as
\eqn\eIVii{\eqalign{
\bhp (u) &= 2\pi \> \zdag_- (\th ) ,
{}~~~~~~~~~~\bhm (u) = 2 \pi \> \zdag_+ (\th )  ~~~~~~~~~~~{\rm for} ~ u>0
\cr
\bhp (u) &= 2\pi i \> Z^+ (\th - i\pi )    ,
{}~~~~~~~~\bhm (u) = 2\pi i \> Z^- (\th - i\pi )    ~~~~~~~~~{\rm for} ~ u<0
.\cr }}
Then the temporal quantization expansion \eIIxvib\ may be written as
\eqn\eIIxx{
\Psi^\pm = \col{\psib^\pm}{\psi^\pm} = \pm \sqm
\int_{-\infty}^\infty \du \> \bh^\pm (u) ~
\col{1/ \squ}{-i\squ} ~ \ez , }

Let us now define the following prescription for analytic continuation
of the $u$-integral in \eIIxx:
\eqn\epercont{\eqalign{
\int_{0}^\infty \du ~ \bhpm (u) &\rightarrow
 \oint_{\CC_<} \dua \( \bpm_< (u) + \bbpm_< (u) \)  \cr
\int_{-\infty}^0 \du ~ \bhpm (u) &\rightarrow
 \int_{\CC_> } \dua \( \bpm_> (u) + \bbpm_> (u) \)  , \cr}}
where
the contour $\CC_<$ is defined to be a closed contour on
the unit circle in the complex $u$-plane,
whereas $\CC_>$
 runs from $0$ to $\infty$ along a ray at an angle $\vphi$
above the negative $x$-axis in the $u$ plane.
The operators on the RHS are defined to have the following
expansions:
\eqn\eperiodic{\eqalign{
\bpm_< (u) &= \pm i \sum_{\om \leq -1/2} \Ga (\ha - \om ) ~ \ma^\om ~
b^\pm_\om u^\om
, ~~~~~\bpm_> (u) = \pm  \sum_{\om \geq 1/2} \frac{2\pi (-1)^{\om + 1/2}}
{\Ga (\om + \ha ) }
 ~ \ma^\om ~ b^\pm_\om u^\om  \cr
\bbpm_< (u) &= \pm  \sum_{\om \leq -1/2} \Ga (\ha - \om ) ~\ma^\om ~ \bbpm_\om
u^{-\om} , ~~~~~
\bbpm_> (u) = \pm i \sum_{\om \geq 1/2} \frac{2\pi (-1)^{\om - 1/2}}
{\Ga (\om + \ha ) }
 ~ \ma^\om ~ \bb^\pm_\om u^{-\om } . \cr}}
One can show, using well-known integral representations for the
Bessel functions, that the analytic continuation \epercont, \eperiodic\
of \eIIxx\ yields the expansion \eIIIxvii\ of radial quantization.
(See \lec.)

Now, consider the states
\eqn\eIVvi{
\dstate = \inv{(2\pi i)^n} \>
\lvac \bh^{\ep_1} (e^{-i\pi} u_1 ) \cdots \bh^{\ep_n} (e^{-i\pi} u_n )
, ~~~~u_i < 0 . }
Using the analytic continuation \epercont, one can map the above
state into the space of radial quantization.  Bearing in mind that
$u_i <0$ in \eIVvi, one uses the second formula in \epercont\ to
obtain the equation \eIIix, where the vacuum and vertex operators
are:
\eqn\eIVvii{\eqalign{
\lva{\Omega} &= \lvac \cr
V^{\ep} (\th ) &= \inv{2\pi i} \( b^\ep_> (e^{-i\pi} u ) + \bb^\ep_>
(e^{-i\pi} u ) \)  . \cr }}

One can easily check the validity of \eIVvii\ by verifying that it
gives the correct form factors in some simple cases.  As explained
above, any field $\Phi$ in the periodic sector corresponds to a
state $\va{\Phi}$ in the space $\hf^{(p)}$ described in section 3, and
the form-factor is computed as
\eqn\eIVviii{
\dstate \Phi \rangle = \lvac V^{\ep_1} (\th_1 ) \cdots V^{\ep_n} (\th_n )
\va{\Phi } . }
Note that since only the radial annihilation operators appear in
\eIVvii, all of the form factors in the periodic sector have the
`free field' property: for a given field, the $n$-particle form factors
are all zero except for $n=n_{\psi} $, where $n_\psi$ is the number
of free fermion fields needed to construct the field.
Using \eIVviii, \eIVvii, and \eIIIxiv\ one finds
\eqn\eIVix{\eqalign{
{}^{\pm} \lva {\th}  \psi^{\mp} (0) \rvac &= \pm \sqrt{\ma u} ,
{}~~~~~~{}^\pm \lva {\th } \psib^\mp (0) \rvac = \pm i \sqrt{\ma /u} \cr
\tstate J_z (0) \rvac &= -\ma (u_1 u_2 )^{1/2}  , ~~~~~
\tstate J_\zbar (0) \rvac = \ma (u_1 u_2 )^{-1/2}  \cr
\tstate T_{zz} (0) \rvac &= \frac{\ma^2}{2}
\( (u_1)^{1/2} (u_2)^{3/2} - (u_1)^{3/2} (u_2)^{1/2}  \)
\cr
\tstate T_{\zbar\zbar} (0) \rvac &= \frac{\ma^2}{2}
\( (u_1)^{-3/2} (u_2)^{-1/2} - (u_1)^{-1/2} (u_2)^{-3/2}  \)
\cr
\tstate T_{z \zbar} (0) \rvac &= \frac{\ma^2}{2}
\(  \( \frac{u_1}{u_2} \)^{1/2} - \( \frac{u_2}{u_1} \)^{1/2}
\)
. \cr }}
All the higher multiparticle form factors are zero for these
fields, since they are all fermion bilinears.
These form factors agree with expressions derived by the standard
methods in the space of particles $\hp$.

\medskip\noindent
4.2 {\it Anti-Periodic Sector}

In this sector, the analytic continuation of \eIIxx\ that reproduces
the radial expansion \eIIIxvii\ is the following:
\eqn\eIIIxiii{
\int_{-\infty}^\infty \du ~ \bhpm (u)
\rightarrow
\(
\int_{\cL} \dua ~ b_\pm (u) ~~ + ~~
\int_{\cR} \dua ~ \bb_\pm (u) \), }
where
$\cL , \cR$ are contours depending on the angular direction $\vphi$ of
the cut displayed in figures 1,2,  and
\midinsert
\epsfxsize = 2in
\vbox{\vskip -.1in\hbox{\centerline{\epsffile{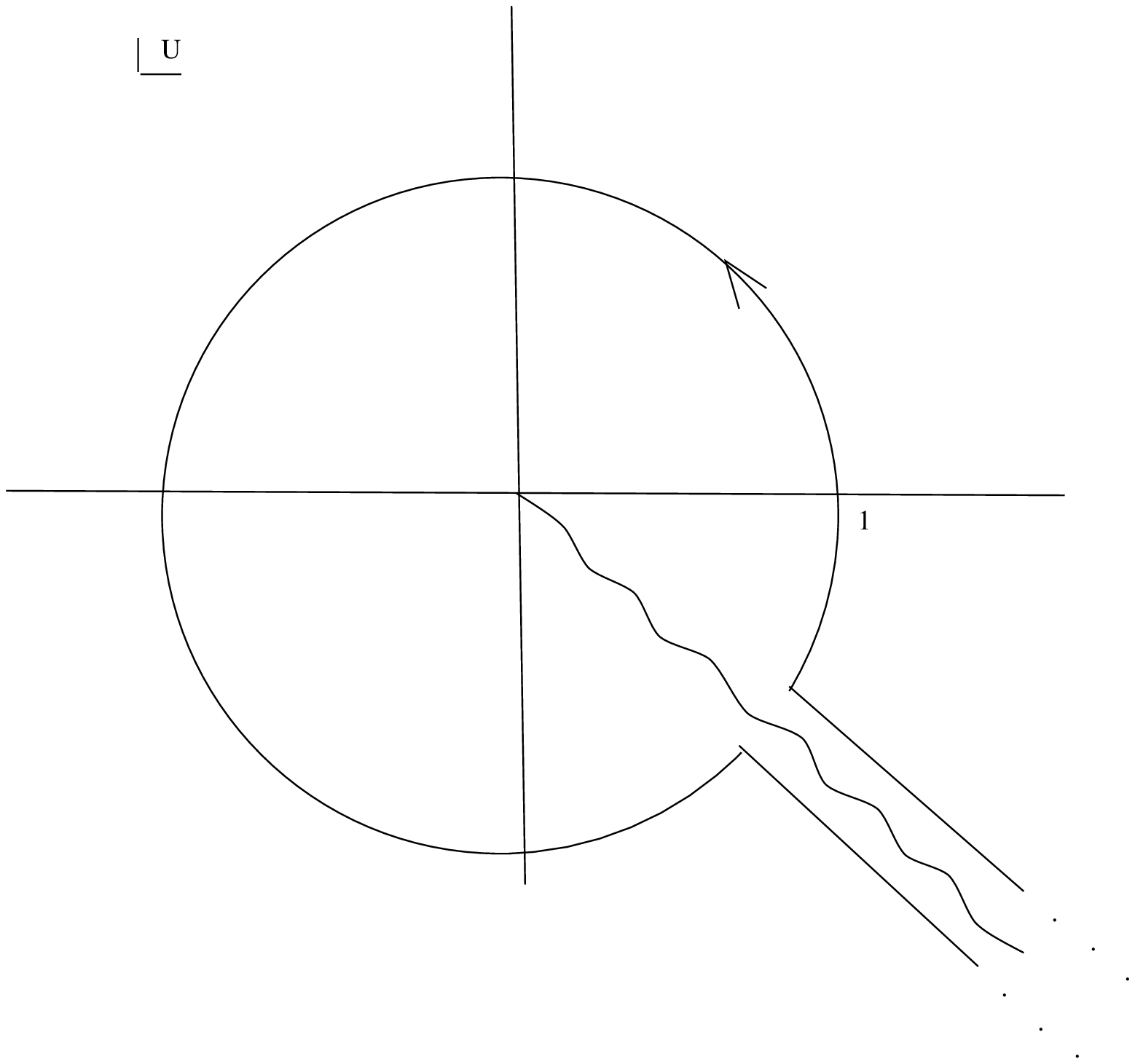}}}
\vskip .1in
{\leftskip .5in \rightskip .5in \noindent \ninerm \baselineskip=10pt
\bigskip\bigskip
Figure 1.
The contour $\CC^L_\vphi$.  The cut (wavy line) is oriented at an
angle $\vphi$ from the negative $y$-axis. The circle is at $|u| = 1$.
\smallskip}} \bigskip
\endinsert
\midinsert
\epsfxsize = 2in
\bigskip
\vbox{\vskip -.1in\hbox{\centerline{\epsffile{contourR.ps}}}
\vskip .1in
{\leftskip .5in \rightskip .5in \noindent \ninerm \baselineskip=10pt
Figure 2.
The contour $\CC^R_\vphi$.  The cut  is oriented at an
angle $\vphi$ from the positive $y$-axis.  The circle is at
$|u| = 1$.
\smallskip}} \bigskip
\endinsert
\eqn\eIIIxiv{\eqalign{
\bpm (u) &= \pm i \sum_{\om \in \Zmath } \Gamma (\ha - \om )
\> \ma^\om ~ b^\pm_\om ~ u^\om   \cr
\bbpm (u) &= \pm  \sum_{\om \in \Zmath } \Gamma (\ha - \om )
\> \ma^\om ~ \bb^\pm_\om ~ u^{-\om}  .    \cr }}

The most significant difference between the anti-periodic
and periodic sectors is that here the analytic continuation
\eIIIxiii\ does not separate the radial creation operators from
the annihilation operators, unlike in \epercont.  This simple
fact is what is responsible for the non-free properties of the
form factors in this sector.

Repeating the reasoning given above for the periodic sector, we
propose that again the formula \eIIix\ is valid, where now
\eqn\eIVxii{\eqalign{
\lva{\Omega}  &= \lva{\ha} + \lva{-\ha}  \cr
V^\ep (\th )  &=  \inv{\sqrt{2\pi^2 i } }
\( b^\ep (e^{-i\pi} u )  + \bb^\ep ( e^{-i\pi} u ) \) . \cr}}
The normalization of the vertex operators was chosen to satisfy the
residue property
\eqn\eIVxiii{
V^+ (\th ) V^- (\th + \beta + i\pi ) ~\sim ~  \inv{i\pi \> \beta} , }
which leads to the proper residue axiom for the multiparticle
form factors.  From \eIIIxvi, one sees that the choice \eIVxii\ for
$\lva{\Omega}$ is equivalent to  the following vacuum expectation values:
\eqn\eIVxiv{
\lvac ~ e^{\pm i \phi (0)/2} ~ \rvac = (c\ma )^{1/4} ~
\lva{\Omega} \pm \ha \rangle = (c\ma )^{1/4} . }

One can use the above construction to compute the form factors
of the fields $\exp (\pm i \phi /2 )$\lec.  For these fields,
all of the form factors with a $U(1)$ neutral combination of an
even number of particles is non-zero.  The result is
\eqn\effii{\eqalign{
& {}^{+++...---...} \lva{\th_1 , \th_2 , \cdots , \th_{2n} }
{}~ e^{\pm i \phi (0) /2 } \rvac \cr
&~~~~=
(c\ma )^{1/4}~    \langle \mp \ha \vert
V^+ (u_1 ) \cdots V^+ (u_n ) V^- (u_{n+1} ) \cdots V^- (u_{2n} )
\vert \pm \ha \rangle \cr
&~~~~= (c\ma )^{1/4} \frac{ (\pm 1)^n }{( i\pi )^n } (-1)^{n(n-1)/2}
\sqrt{u_1 \cdots u_{2n} }
\( \prod_{i=1}^n \( \frac{u_{i+n}}{u_i} \)^{\pm 1/2} \)
\( \prod_{i<j \leq n} (u_i - u_j ) \)  \cr
& ~~~~~~~~\times \( \prod_{n+1 \leq i < j } (u_i - u_j ) \)
\( \prod_{r=1}^n \prod_{s=n+1}^{2n} \inv{u_r + u_s } \)   . \cr } }
The above computation can be done using the Wick theorem with the
2-point functions
\eqn\eIVvii{\eqalign{
{}_L \lvacm ~ b^+ (u) \> b^- (u' ) ~ \rvacp_L
= {}_R \lvacp ~ \bb^+ (u) \> \bb^- (u' ) ~ \rvacm_R  &= \pi
\frac{u'}{u+u'} \cr
{}_L \lvacp ~ b^+ (u) \> b^- (u' ) ~ \rvacm_L
= {}_R \lvacm ~ \bb^+ (u) \> \bb^- (u' ) ~ \rvacp_R  &= - \pi
\frac{u}{u+u'} . \cr}}
However the computation is more easily
done using the bosonization techniques of
the next section.
After some algebraic manipulation,
one can see that these expressions agree with the known results,
though they were
originally computed using rather different
methods\ref\rmss{\MSS}\rform\foot{The overall numerical factors in
\effii\ differ from the ones in \rform\ however they agree with
results implicit in \rmss. Here the correct normalization is fixed by
the residue property \eIVxiii. }.

\newsec{Bosonization in Momentum Space}

Bosonization is a well-known construction in position space,
and finds its most precise statement in conformal field theory.
In this section, we will construct an exact bosonization in
{\it momentum space} for the massive theory we are considering.

Though position space conformal bosonization is not physically
relevant here, we review it in order to simply present a
mathematical result that will facilitate understanding of the
results presented below.  Given a set of fermionic oscillators
satisfying $\{ c^+_\om , c^-_{\om'} \} = \delta_{\om + \om' , 0} $,
define a bosonic operator as follows:
\eqn\eVi{
\rho_n = \ma^n \sum_\om : c^+_{n-\om} c^-_\om :  }
These operators satisfy an infinite Heisenberg algebra:
\eqn\eVii{
[\rho_n , \rho_{n'} ] = n \delta_{n+n' ,0}  ~. }
Bosonization amounts to using the $\rho_n$ to construct
a generating function for the $c^\pm_\om $ as follows.
Define a field
\eqn\eViii{
-i \rho (z) = \sum_{n\neq 0} \rho_n \> \frac{z^{-n}}{n}
-\rho_0 \log (z) - \tilde{\rho}_0  }
where $\rt_0$ is defined to satisfy
\eqn\eViv{
[\rho_0 , \rt_0 ] = 1. }
Then one has
\eqn\eVv{
c^\pm (z) = \sum_\om c^\pm_\om \> z^{-\om - 1/2}
= : e^{\pm i \rho (z) } : }
In conformal field theory, the above statements correspond to
the exact bosonization of a free massless fermion, where $z$ is
the left-moving light-cone coordinate. (See e.g. \ref\rgins{\Ginsparg}.)
We now turn to the massive theory.

\medskip\noindent
5.1 {\it Anti-Periodic Sector}

As shown in \lec, in the massive theory one can use the constants of
motion to formulate an exact bosonization.  We review this here in
order to compare with the periodic sector.
In radial quantization, define the inner product of two spinors
$A = \col{\bar{a}}{a}$, $B = \col{\bar{b}}{b}$ as
\eqn\eVii{
(A,B) = \inv{4\pi} \int_{-\pi}^\pi ~ r d\vphi
\( e^{ i\vphi} ~ a \>  b ~+~ e^{-i\vphi} ~ \bar{a} \> \bar{b} \) . }
The conserved charges constructed in
\ref\raass{\AASS}\lec\  can all be expressed using the above inner
product:
\eqn\eVvb{\eqalign{
Q^\pm_{-n} &= \ha   \( \Psi^\pm , \d^n_z \Psi^\pm \) , ~~~~~
Q^\pm_{n} = \ha  \( \Psi^\pm , \d^n_\zb \Psi^\pm \) \cr
\al_{-n} &= {(-)^{n}}  \( \Psi^+ , \d^n_z \Psi^- \) , ~~~~~
\al_{n} = {(-)^{n}} :\( \Psi^+ , \d^n_\zb \Psi^- \) : ~
, \cr }}
where $n\geq 0$ is an integer, and for $Q^\pm_n $, $n$ is odd.
The operators $\left\{ Q^\pm_n , \al_ m , ~m ~ {\rm even} \right\}$
generate a twisted affine $\slh$ algebra at level $k=0$, and
the $\al_n$ for $n ~ {\rm odd}$ are the usual infinity of
integrals of motion at odd integer Lorentz spin.  In radial
quantization, using Wronskian identities for the Bessel functions,
one finds that the above conserved charges split into Left
and Right pieces:
\eqn\eVvii{
Q^\pm_n ~=~ Q^{\pm , L}_n ~+~ Q^{\pm ,R}_{-n}  ,~~~~~~~
\al_n ~=~ \al^L_n ~+~ \al^R_{-n} .}
Both the L and the R pieces of the $\slh$ algebra separately
satisfy a level $k=1$ algebra in the anti-periodic sector, and
a level $k=0$ algebra in the periodic sector\lec.  Here we are
mainly interested in the $\al^{L,R}_n$ which have the
explicit expressions:
\eqn\eVviii{
\al^L_n =
 {\ma^{|n| + n}} ~ \sum_{\om \in S^{(a,p)}_n }
\frac{\Ga (\ha + \om -n )}{\Ga (\ha + \om )} ~
 : b^+_{n-\om} ~ b^-_{\om } :  }
where the sums over $\om$ run over $S^{(p)}_n$ ($S^{(a)}_n $) for
the periodic (anti-periodic) sector, and
\eqn\esn{
S^{(a)}_n = \Zmath , ~~~\forall ~ n,  ~~~~~
S^{(p)}_n = \left\{ \om \in \Zmath + 1/2 : |\om - n/2| > |n|/2 \right\} . }
Identical expressions apply to $\al^{R}_n $ with
$b^\pm_\om \to \bb^\pm_\om $.

In the anti-periodic sector, since $\al^{L,R}_n$ satisfy two separate
Heisenberg algebras, they can be used to construct a bosonization.
Define
$$\rho_n = \ma^{-|n|} \al^{L}_n , ~~~~~~\rb_n = \ma^{-|n|} \al^R_n  ,$$
satisfying \eVii\ and $[\rb_n , \rb_{n'} ] = n \delta_{n+n',0}$,
and define the momentum space fields (recall $u= e^\th$):
\eqn\eVIx{\eqalign{
-i \rho (u) &=    \sum_{n\neq 0}  ~ \rho_n  ~ \frac{u^n}{n}
  + \rho_0 \log (u) - \rt_0  \cr
-i \rb  (u) &=    \sum_{n\neq 0}  ~\rb_n  ~ \frac{u^{-n}}{n}
   - \rb_0 \log (u) - \rbt_0  , \cr }}
where one also has $[\rb_0 , \rbt_0 ] = 1$.
We further define an auxiliary vacuum satisfying
\eqn\avac{
\al_n^L \emrvac = \al_n^R \emrvac = 0 , ~~n\geq 0; ~~~~~
\alt^L_0 \emrvac ~, ~~\alt^R_0 \emrvac \neq 0 . }
This vacuum $\emrvac$ is not to be confused with the physical
vacuum $\rvac$ which resides in the periodic sector.  One has
the following  expectation values:
\eqn\eVIxiv{\eqalign{
 \emlvac ~ \rho (u) ~ \rho (u') ~\emrvac
&= -\log ( 1/u - 1/u' ) \cr
 \emlvac ~ \rb (u) ~ \rb (u') ~\emrvac  &= -\log (u - u' ) \cr}}
\medskip
\eqn\eVIxv{\eqalign{
 \emlvac \prod_i ~ e^{i\al_i \rho (u_i )} ~ \emrvac
&= \prod_{i< j} \( 1/u_i - 1/u_j \)^{\al_i \al_j }  \cr
 \emlvac \prod_i ~ e^{i\al_i \rb (u_i )} ~ \emrvac
&= \prod_{i< j} \( u_i - u_j \)^{\al_i \al_j }  . \cr}}
The bosonized expressions for the operators
$b^\pm (u) , \bb^\pm (u)$ and the states $\va{\pm \ha}$
follow from the basic commutation relations
\eqn\eVxvi{\eqalign{
\[ \al_n^L , b^\pm (u) \] &= (\pm 1)^{n+1} \ma^{|n|} u^{-n} ~ b^\pm (u) \cr
\[ \al_n^R , \bb^\pm (u) \] &= (\pm 1)^{n+1}
\ma^{|n|} u^{n} ~ \bb^\pm (u) \cr }}
and the 2-point functions \eIVvii.  The commutation relations
\eVxvi\ are fundamental in the sense that they describe how the
conserved charges are represented on asymptotic multiparticle
states.  One finds
\eqn\eVvii{
\inv{\sqrt{\pm \pi u}}
{}~ b^\pm (u)   = : e^{\pm i \rho (\pm u)} :
, ~~~~~~
\frac{\pm 1}{\sqrt{\pm \pi u }}
{}~  \bb^\pm (u)  = : e^{\pm i \rb (\pm u )} :
}
where $-u = e^{-i\pi} u $, and
\eqn\eVIxvi{\eqalign{
\va{\pm \ha}_L ~ &= ~ : e^{\pm i \rho (\infty )/2 } :  \emrvac_L ,
{}~~~~~~~~~~~~~
\va{\pm \ha}_R ~ = ~: e^{\pm i \rb (0) /2 }:  \emrvac_R \cr
{}_L
\lva{\pm \ha} &= \lim_{u\to 0} ~ u^{-1/4} ~ \emlvac :
e^{\pm i \rho (u)/2 } :
, ~~~~~
{}_R
\lva{\pm\ha} = \lim_{u\to \infty} ~ u^{1/4} ~ \emlvac
: e^{\pm i \rb (u) /2 } :
.\cr}}
One can easily check that this construction reproduces the
form factors \effii.

The above bosonization is mathematically identical to the one
described in \eVi - \eVv. One can see this explicitly by
making the canonical transformation (by `canonical', we
mean one that preserves the commutation relations):
\eqn\eVix{
c^\pm_\om = \frac{(\pm 1)^\om }{\sqrt{\pi}}  \Ga (\ha - \om ) ~ b^\pm_\om
, ~~~~~
\bar{c}^\pm_\om = \frac{(\pm 1)^\om }{\sqrt{\pi}}  \Ga (\ha - \om )
{}~ \bb^\pm_\om  . }

\medskip\noindent
5.2 {\it Periodic Sector}

In the periodic sector, one cannot use the constants of the motion
$\al^{L,R}_n$ to construct a bosonization, due to the fact that in
this sector: $[ \al^L_n , \al^R_m ] = [\al^R_n , \al^R_m ] = 0$.
This can be traced to the gaps in the sets $S^{(p)}_n$.
As explained in \lec, this had to be the case for the following
reason:  the physical vacuum $\rvac$ is only invariant with respect
to the conserved charges $\al_n$ if $\al^L_n $ and $\al^R_n$ separately
commute.

One can still construct a bosonization in this sector modeled after the
formulas \eVi-\eVv. Namely define  the canonical transformation
\eqn\eVx{\eqalign{
c^\pm_\om &= \frac{\Ga (\ha - \om )}{\sqrt{2\pi}}  \> b^\pm_\om  , ~~~~~
\bar{c}^\pm_\om = \frac{\Ga (\ha  - \om )}{\sqrt{2\pi}} \> \bb^\pm_\om
{}~~~~~{\rm for} ~ \om \leq -1/2 \cr
c^\pm_\om &= \frac{\sqrt{2\pi}}{ \Ga (\ha + \om )}  \> b^\pm_\om  , ~~~~~
\bar{c}^\pm_\om = \frac{\sqrt{2\pi}}{\Ga (\ha  - \om )} \> \bb^\pm_\om
{}~~~~~{\rm for} ~ \om \geq 1/2 , \cr }}
and define
$\rho_n , \rb_n$ as in \eVi, and $\rho (u) ,  \rb (u)$ as in
\eVIx.  Then one has the bosonization formulas
\eqn\eVxi{\eqalign{
\pm \inv{\sqrt{2\pi}}
\( \sqrt{-u} \> b^\pm_> (-u)  - i \sqrt{u} \> b^\pm_< (u) \)
&= : e^{\pm i \rho (u)} : \cr
\pm \frac{i} {\sqrt{2\pi}}
\( \inv{\sqrt{-u}} \> \bb^\pm_> (-u)  - \frac{i}{\sqrt{u}} \> b^\pm_< (u) \)
&= : e^{\pm i \rb (u)} : . \cr}}

\newsec{Conclusions}

Though we have limited ourselves to perhaps the simplest possible
case of the free-fermion point of the sine-Gordon theory, we
believe the ideas presented here can lead to a new framework for
computing form factors in massive integrable quantum field theory.
In this approach, since a complete description of the space of
fields $\hf$ is provided from the outset via radial quantization,
the complete set of solutions to the form factor bootstrap is
automatically yielded.
In the bosonized construction in section 5, an important role was
played by the affine $\slh$ symmetry.  Since away from the free-fermion
point this symmetry is deformed to a $\CU_q (\slh )$
symmetry\ref\rnlc{\BLnlc}, with $q= \exp (-2\pi i/{\hat{\beta}}^2 )$,
this quantum affine symmetry is expected to be important for
the general construction.  The results contained in
\ref\rfr{\FrResh}\ref\rdefkz{F. A. Smirnov, Int. J. Mod. Phys. A7,
Suppl. 1B (1992) 813.}\ref\rfj{\Frenkeli}\ should prove
useful.

Readers with some familiarity with conformal field theory will
doubtless see the strong parallels of this subject with the
work presented here.  For the example we have developed,
we have shown that in radial quantization
form factors can be computed as correlation functions in momentum space,
and these correlation functions are very similar in structure to
conformal spacetime correlation functions.  Furthermore, for the
purposes of computing form factors, one can describe the space of fields
in the same way as is done in the ultraviolet conformal field theory.
In a definite sense, we have shown that starting from a description
of the space of fields in a conformal field theory and the basic
operators from which one constructs this space (in our case, we mean
the operators $b^\pm_\om , \bb^\pm_\om $), then one can reconstruct
a massive theory and its form factors by constructing the vertex operators.
It is important to understand if this is possible more generally.

\bigskip
\centerline{Acknowledgements}

We would like to thank Denis Bernard  and Sergei Lukyanov
for discussions.
This work is supported
by an  Alfred P. Sloan Foundation fellowship,  and the
National Science Foundation in part through the
National Young Investigator program.

\vfill\eject

\listrefs
\end